\input harvmac

\def\inbar{\vrule height1.5ex width.4pt depth0pt}
\def\IC{{\relax\,\hbox{$\inbar\kern-.3em{\rm C}$}}}
\def\IR{{\relax\,\hbox{$\inbar\kern-.3em{\rm R}$}}}
\def\IP{{\relax\,\hbox{$\inbar\kern-.3em{\rm P}$}}}
\def\Zbar{{\bar{Z}}}
\def\lambdabar{{\bar{\lambda}}}
\def\hap{\textstyle{p \over 2}}
\def\haq{\textstyle{q \over 2}}
\def\hapone{\textstyle{p_1 \over 2}}
\def\haptwo{\textstyle{p_2 \over 2}}
\def\p{\partial}
\def\pbar{{\bar{\partial}}}
\def\zbar{{\bar{z}}}

\lref\BerensteinJQone{
  D.~Berenstein, J.~M.~Maldacena and H.~Nastase,
  ``Strings in flat space and pp waves from N = 4 super Yang Mills,''
  JHEP {\bf 0204}, 013 (2002)
  [arXiv:hep-th/0202021].
}

\lref\PlefkaBK{
  J.~Plefka,
  ``Spinning strings and integrable spin chains in the AdS/CFT
  correspondence,''
  arXiv:hep-th/0507136.
}

\lref\TseytlinII{
  A.~A.~Tseytlin,
  ``Spinning strings and AdS/CFT duality,''
  arXiv:hep-th/0311139.
}

\lref\TseytlinXA{
  A.~A.~Tseytlin,
  ``Semiclassical strings and AdS/CFT,''
  arXiv:hep-th/0409296.
}

\lref\ZaremboHP{
  K.~Zarembo,
  ``Semiclassical Bethe ansatz and AdS/CFT,''
  Comptes Rendus Physique {\bf 5}, 1081 (2004)
  [Fortsch.\ Phys.\  {\bf 53}, 647 (2005)]
  [arXiv:hep-th/0411191].
}

\lref\MandalFS{
  G.~Mandal, N.~V.~Suryanarayana and S.~R.~Wadia,
  ``Aspects of semiclassical strings in $AdS_{5}$,''
  Phys.\ Lett.\  B {\bf 543}, 81 (2002)
  [arXiv:hep-th/0206103].
}

\lref\DolanPS{
  L.~Dolan, C.~R.~Nappi and E.~Witten,
  ``Yangian symmetry in D = 4 superconformal Yang-Mills theory,''
  arXiv:hep-th/0401243.
}

\lref\DoreyXN{
  N.~Dorey, D.~M.~Hofman and J.~Maldacena,
  ``On the singularities of the magnon S-matrix,''
  arXiv:hep-th/0703104.
}

\lref\EichenherrHZ{
  H.~Eichenherr and M.~Forger,
  ``More About Nonlinear Sigma Models On Symmetric Spaces,''
  Nucl.\ Phys.\ B {\bf 164}, 528 (1980)
  [Erratum-ibid.\ B {\bf 282}, 745 (1987)].
}

\lref\StaudacherTK{
  M.~Staudacher,
  ``The factorized S-matrix of CFT/AdS,''
  JHEP {\bf 0505}, 054 (2005)
  [arXiv:hep-th/0412188].
}

\lref\MinahanVE{
  J.~A.~Minahan and K.~Zarembo,
  ``The Bethe-ansatz for N = 4 super Yang-Mills,''
  JHEP {\bf 0303}, 013 (2003)
  [arXiv:hep-th/0212208].
}

\lref\JanikDC{
  R.~A.~Janik,
  ``The AdS${}_5 \times S^5$
  superstring worldsheet S-matrix and crossing symmetry,''
  Phys.\ Rev.\ D {\bf 73}, 086006 (2006)
  [arXiv:hep-th/0603038].
}

\lref\BeisertYB{
  N.~Beisert and M.~Staudacher,
  ``The N = 4 SYM integrable super spin chain,''
  Nucl.\ Phys.\ B {\bf 670}, 439 (2003)
  [arXiv:hep-th/0307042].
}

\lref\BenaWD{
  I.~Bena, J.~Polchinski and R.~Roiban,
  ``Hidden symmetries of the AdS$_5 \times S^5$ superstring,''
  Phys.\ Rev.\ D {\bf 69}, 046002 (2004)
  [arXiv:hep-th/0305116].
}

\lref\BeisertRY{
  N.~Beisert,
  ``The dilatation operator of N = 4 super Yang-Mills theory and
  integrability,''
  Phys.\ Rept.\  {\bf 405}, 1 (2005)
  [arXiv:hep-th/0407277].
}

\lref\BakasBM{
  I.~Bakas, Q.~H.~Park and H.~J.~Shin,
  ``Lagrangian Formulation of Symmetric Space sine-Gordon Models,''
  Phys.\ Lett.\ B {\bf 372}, 45 (1996)
  [arXiv:hep-th/9512030].
}

\lref\PohlmeyerCH{
  K.~Pohlmeyer and K.~H.~Rehren,
  ``Reduction Of The Two-Dimensional O(N) Nonlinear Sigma Model,''
  J.\ Math.\ Phys.\  {\bf 20}, 2628 (1979).
}

\lref\SantambrogioSB{
  A.~Santambrogio and D.~Zanon,
  ``Exact anomalous dimensions of N = 4 Yang-Mills operators with large
  R charge,''
  Phys.\ Lett.\ B {\bf 545}, 425 (2002)
  [arXiv:hep-th/0206079].
}

\lref\BerensteinJQtwo{
  D.~Berenstein, D.~H.~Correa and S.~E.~Vazquez,
  ``All loop BMN state energies from matrices,''
  JHEP {\bf 0602}, 048 (2006)
  [arXiv:hep-th/0509015].
}

\lref\HofmanXT{
  D.~M.~Hofman and J.~M.~Maldacena,
  ``Giant magnons,''
  arXiv:hep-th/0604135.
}

\lref\HofmanXT{
  D.~M.~Hofman and J.~M.~Maldacena,
  ``Giant magnons,''
  J.\ Phys.\ A  {\bf 39}, 13095 (2006)
  [arXiv:hep-th/0604135].
}

\lref\BeisertTM{
  N.~Beisert,
  ``The su(2$|$2) dynamic S-matrix,''
  arXiv:hep-th/0511082.
}

\lref\ArutyunovVX{
  G.~Arutyunov, S.~Frolov and M.~Staudacher,
  ``Bethe ansatz for quantum strings,''
  JHEP {\bf 0410}, 016 (2004)
  [arXiv:hep-th/0406256].
}

\lref\MannJR{
  N.~Mann and J.~Polchinski,
  ``Finite density states in integrable conformal field theories,''
  arXiv:hep-th/0408162.
}

\lref\MannAB{
  N.~Mann and J.~Polchinski,
  ``Bethe ansatz for a quantum supercoset sigma model,''
  Phys.\ Rev.\ D {\bf 72}, 086002 (2005)
  [arXiv:hep-th/0508232].
}

\lref\PohlmeyerNB{
  K.~Pohlmeyer,
  ``Integrable Hamiltonian Systems And Interactions Through Quadratic
  Constraints,''
  Commun.\ Math.\ Phys.\  {\bf 46}, 207 (1976).
}

\lref\MikhailovQV{
  A.~Mikhailov,
  ``An action variable of the sine-Gordon model,''
  arXiv:hep-th/0504035.
}

\lref\MikhailovZD{
  A.~Mikhailov,
  ``B\"acklund transformations, energy shift and the plane wave limit,''
  arXiv:hep-th/0507261.
}

\lref\MikhailovSY{
  A.~Mikhailov,
  ``A nonlocal Poisson bracket of the sine-Gordon model,''
  arXiv:hep-th/0511069.
}

\lref\ArutyunovGS{
  G.~Arutyunov, S.~Frolov and M.~Zamaklar,
  ``Finite-size effects from giant magnons,''
  arXiv:hep-th/0606126.
}

\lref\MinahanBD{
  J.~A.~Minahan, A.~Tirziu and A.~A.~Tseytlin,
  ``Infinite spin limit of semiclassical string states,''
  JHEP {\bf 0608}, 049 (2006)
  [arXiv:hep-th/0606145].
}

\lref\ChuAE{
  C.~S.~Chu, G.~Georgiou and V.~V.~Khoze,
  ``Magnons, classical strings and beta-deformations,''
  JHEP {\bf 0611}, 093 (2006)
  [arXiv:hep-th/0606220].
}

\lref\ChenGE{
  H.~Y.~Chen, N.~Dorey and K.~Okamura,
  ``Dyonic giant magnons,''
  JHEP {\bf 0609}, 024 (2006)
  [arXiv:hep-th/0605155].
}

\lref\DoreyDQ{
  N.~Dorey,
  ``Magnon bound states and the AdS/CFT correspondence,''
  J.\ Phys.\ A  {\bf 39}, 13119 (2006)
  [arXiv:hep-th/0604175].
}

\lref\ZakharovPP{
  V.~E.~Zakharov and A.~V.~Mikhailov,
  ``Relativistically Invariant Two-Dimensional Models In Field Theory
  Integrable By The Inverse Problem Technique. (In Russian),''
  Sov.\ Phys.\ JETP {\bf 47}, 1017 (1978)
  [Zh.\ Eksp.\ Teor.\ Fiz.\  {\bf 74}, 1953 (1978)].
}

\lref\ZakharovTY{
  V.~E.~Zakharov and A.~V.~Mikhailov,
  ``On The Integrability Of Classical Spinor Models In Two-Dimensional
  Space-Time,''
  Commun.\ Math.\ Phys.\  {\bf 74}, 21 (1980).
}

\lref\HarnadWE{
  J.~P.~Harnad, Y.~Saint Aubin and S.~Shnider,
  ``Backlund Transformations For Nonlinear Sigma Models With Values In
  Riemannian Symmetric Spaces,''
  Commun.\ Math.\ Phys.\  {\bf 92}, 329 (1984).
}

\lref\SaintAubinAN{
  Y.~Saint Aubin,
  ``B\"acklund Transformations And Soliton Type Solutions For Sigma Models With
  Values In Real Grassmannian Spaces,''
  Lett.\ Math.\ Phys.\  {\bf 6}, 441 (1982).
}

\lref\ArutyunovUJ{
  G.~Arutyunov, S.~Frolov, J.~Russo and A.~A.~Tseytlin,
  ``Spinning strings in AdS${}_5 \times S^5$ and integrable systems,''
  Nucl.\ Phys.\ B {\bf 671}, 3 (2003)
  [arXiv:hep-th/0307191].
}

\lref\OgielskiHV{
  A.~T.~Ogielski, M.~K.~Prasad, A.~Sinha and L.~L.~Wang,
  ``Backlund Transformations And Local Conservation Laws For Principal Chiral
  Fields,''
  Phys.\ Lett.\ B {\bf 91}, 387 (1980).
}

\lref\juannote{
  Equivalent solutions have also been obtained by J.~Maldacena
  and A.~Mikhailov
  from  the B\"acklund transformation
  \MikhailovZD\ (private communication).
} 

\Title
{\vbox{
\baselineskip12pt
\hbox{hep-th/0607009}
}}
{\vbox{
\centerline{Dressing the Giant Magnon}
}}

\centerline{
Marcus Spradlin${}^{}$
and Anastasia Volovich${}^{}$
}

\bigskip

\centerline{
{\tt spradlin@ias.edu},
{\tt nastja@ias.edu}
}

\vskip .5in
\centerline{Institute for Advanced Study}
\centerline{Princeton, New Jersey 08540 USA}
\bigskip
\centerline{and}
\bigskip
\centerline{Brown University}
\centerline{Providence, Rhode Island 02912 USA}

\vskip .5in
\centerline{\bf Abstract}

We apply the dressing method
to construct new classical string solutions 
describing various scattering and bound
states of magnons.
These solutions carry one, two or three SO(6) charges and correspond
to multi-soliton configurations in the generalized
sine-Gordon models.

\Date{}

\listtoc
\writetoc

\newsec{Introduction}

Integrability on both sides of AdS/CFT
\refs{\MandalFS,\BenaWD,\DolanPS,\MinahanVE,\BeisertYB}
has enabled many quantitative checks of the correspondence
(see \refs{\TseytlinII,\BeisertRY,\TseytlinXA,\ZaremboHP,\PlefkaBK}
for reviews).
At weak 't Hooft coupling, anomalous dimensions of
gauge theory operators can be calculated using the Bethe ansatz of an
integrable spin chain. At strong coupling, string theory
becomes tractable in the semiclassical limit where one
can study the energies of the corresponding classical string configurations.
Understanding in detail the interpolation between weak and
strong coupling remains an outstanding problem.

Recently Hofman and Maldacena \HofmanXT\
suggested a particular
limit where the spectrum simplifies on both sides of
the correspondence.  They considered operators with infinite
energy $\Delta$ and
U(1) R-charge $J$ but finite $\Delta - J$ and fixed
`worldsheet' momentum $p$.
The simplest nontrivial example of such an operator is
\eqn\hm{
{\cal O}_p \sim \sum_l e^{ipl}\, (\cdots ZZZWZZZ \cdots),
}
where $Z$ is a scalar field with R-charge $J=1$ and $W$ is an excitation
inserted at position $l$ along
the chain.
Note that this is a formal limit where the operator becomes infinitely
long and we thus ignore taking the trace (and hence we ignore the
cyclicity constraint which would normally set the total momentum to zero).

This limit is different from
the BMN limit \BerensteinJQone\ and has the nice feature that it
decouples quantum effects characterized by
the 't Hooft coupling $\lambda$ from finite $J$ effects
\refs{\MannJR,\MannAB,\HofmanXT}.
In this limit the spectrum on
both sides can be analyzed in terms of asymptotic states
and the $S$-matrix describing their scattering (this is the asymptotic
$S$-matrix of \StaudacherTK).
The general state can have any number of elementary magnons as well
as bound states thereof.

Hofman and Maldacena identified
the elementary magnon \hm\ with a particular
string
configuration moving on an $\IR \times S^2$ subspace of
AdS${}_5 \times S^5$, which they called the `giant magnon.'
Classical string theory on $\IR \times S^2$ is equivalent to
classical sine-Gordon theory
\refs{\PohlmeyerNB,\MikhailovQV,\MikhailovZD,\MikhailovSY},
and the giant magnon solution of \HofmanXT\ corresponds to the
sine-Gordon soliton.
Using this map to sine-Gordon theory, the scattering
phase of two magnons was computed and shown to match the
large $\lambda$ limit of the conjecture of \ArutyunovVX.

In \ChenGE\ a solution describing
a giant magnon moving on $\IR \times S^3$ with
two angular momenta was constructed, after the existence of such
a state had been shown, and a particular case considered, in \DoreyDQ.
The two-charge giant magnon has infinite $J$ just like \hm, and in addition
carries some finite amount $J_2$ of angular momentum in an orthogonal plane.
This solution was obtained by exploiting the
correspondence between classical string theory on $\IR \times S^3$
and the complex sine-Gordon model.
In contrast to \hm, it corresponds not to a single excitation $W$
but to a bound state of many such excitations carrying a finite macroscopic
amount of $J_2$ charge.
More recent work on giant magnons has considered finite $J$ effects
\ArutyunovGS, some quantum corrections
\MinahanBD, and giant magnon solutions for $\beta$--deformed
AdS${}_5 \times S^5$ \ChuAE.

The aim of this paper is to lay the foundation for a study of more
general giant magnon solutions on $\IR \times S^5$.
We define a giant magnon to be any open string on $\IR \times S^{N-1}$
whose endpoints move at the speed of light along the equator of the
sphere.  One can build a physical closed string solution
from two or more giant magnons by
attaching the beginning of each giant magnon to the end of another.

Previous studies \refs{\HofmanXT,\ChenGE}
have employed the correspondence between classical string theory on
$\IR \times S^2$ (or $\IR \times S^3$) and the sine-Gordon (or complex
sine-Gordon) model.  More generally, string theory on
$\IR \times S^{N-1}$ is classically equivalent to the so-called SO($N$) SSSG
(symmetric space sine-Gordon) model
\refs{\PohlmeyerCH,\BakasBM}.  An advantage of using the sine-Gordon
formulation of the problem is that explicit formulas are known for
arbitrary $n$-soliton configurations in these theories.
The disadvantage of using the sine-Gordon formulation is that the map
between the sine-Gordon variables and the string sigma-model variables
$X_i$ describing the embedding of the string into $\IR \times S^{N-1}$
is nonlinear and difficult or impossible to invert in practice for any
but the simplest configurations.

In this paper we instead focus directly on the
SO($N$) vector model
describing strings on $\IR \times S^{N-1}$ and the SU(2) principal
chiral model describing strings on $\IR \times S^3$.
Being integrable, there exists a procedure for directly constructing
their soliton solutions.
We employ the dressing method
\refs{\ZakharovPP,\ZakharovTY,
\HarnadWE}
to construct classical string solutions corresponding to various
scattering and bound states of magnons, as well as scattering states
of bound states.\foot{The
previous sentence highlights a possible terminological confusion
in this subject.  In this paper we consider only single giant magnons;
that is, single open strings, corresponding to a single operator
but with a possibly arbitrary number of magnon excitations $W$.
The notion of `soliton number' is well-defined in the integrable
SO($N$) vector model, so we will characterize giant magnon solutions
according to how many solitons they carry.  Each soliton may correspond
to one magnon $W$ or to a bound state of many magnons.}

We use the dressing method to rederive the previously known giant magnon
solutions (5.5) and (4.15),
which both correspond to single sine-Gordon solitons, and further
use it to construct
several new solutions (4.18), (5.10), (5.14), and (5.19)
corresponding to scattering or bound states
of two solitons
carrying one, two or three
SO(6) charges.
(Solutions equivalent to (5.10) and (5.14) have also been obtained
by J.~Maldacena
and A.~Mikhailov
from the B\"acklund transformation \juannote.)
Moreover, as we discuss below,
the dressing method allows general
$n$-soliton scattering and bound states
to be constructed algebraically.

It is an important open problem to determine an overall $\lambda$-dependent
phase factor in the magnon $S$-matrix \refs{\BeisertTM,\JanikDC}, whose
zeros and poles must contain information about
the spectrum of magnon bound states.
We calculate the dispersion relations for all of the solutions
constructed in this paper,
but we do not address here the calculation of the scattering phase.
At the semiclassical level, it can be computed
by simply translating the result from the corresponding sine-Gordon
picture, as was done in \HofmanXT\ for two magnons on $\IR \times S^2$.
The calculation of quantum corrections to the scattering phase would
require the explicit formulas presented in (4.18), (5.10) or (5.19)
below since
the correspondence with the sine-Gordon
model breaks down at the quantum level.

We begin in section 2 with a brief statement of 
our notation and the equations to be
solved.
In section 3 we review the dressing method for the principal chiral model,
and explain how to apply it to the SO($N$) vector model by a particular
embedding.
In sections 4 and 5 we apply this method to construct explicit multi-soliton
string configurations for $\IR \times S^3$ and $\IR \times S^{N-1}$
respectively.

\newsec{Giant Magnon Preliminaries}

We use worldsheet coordinates $t$ (identified with physical time)
and $x$, which ranges from $-\infty$ to $+\infty$.
In conformal gauge, a giant magnon is a solution of the sigma model
equations of motion
(we use $z = \ha (x-t)$, $\zbar = \ha (x+t)$)
\eqn\stringeom{
\pbar \p X_i + (\p X_j \pbar X_j) X_i = 0,
\qquad
X_i X_i = 1,
}
subject to the Virasoro constraints
\eqn\virasoro{
\p X_i \p X_i = \pbar X_i  \pbar X_i = 1.
}
When useful, we will employ the complex coordinates
\eqn\complexZ{
Z_1 = X_1 + i X_2, \qquad
Z_2 = X_3 + i X_4, \qquad
Z_3 = X_5 + i X_6.
}
The boundary conditions for a giant magnon at fixed $t$ are
\eqn\magnondef{
\eqalign{
Z_1(t, x \to \pm \infty) &= e^{i t \pm i p/2 + i \alpha}, \cr
Z_i(t, x \to \pm \infty) &= 0, \qquad i = 2,3,
}}
where $\alpha$ is any real constant and $p$ represents the total
worldsheet momentum of the magnon.
Geometrically, $p$ represents the difference
in longitude between the two endpoints of
the string on the equator of the $S^5$.
The first condition \magnondef\ only defines $p$ modulo $2 \pi$.
Although this is sufficient for giant magnon states corresponding
to a single soliton, more general giant magnons
corresponding to scattering or bound states of many solitons can
carry arbitrary $p$.
We can define the total momentum
\eqn\betterp{
p = {1 \over i} \int_{-\infty}^{+\infty} dx 
\,
{d \over dx} \log Z_1.
}
In addition to $p$, giant magnons can be characterized by
the conserved charges
\eqn\charges{\eqalign{
\Delta - J &= {\sqrt{\lambda} \over 2 \pi}
\int_{-\infty}^{+\infty} dx
\, \left( 1 - {\rm Im}[\Zbar_1 \p_t Z_1] \right),\cr
J_i &= {\sqrt{\lambda} \over 2 \pi}
\int_{-\infty}^{+\infty} dx \, {\rm Im} [\Zbar_i \p_t Z_i], \qquad
i=2,3,
}}
where $\lambda$ is the 't Hooft coupling.
Note that $\Delta$ and $J$ are separately infinite for a giant
magnon; only their difference
is finite.

\newsec{Review of the Dressing Method}

In this section we briefly review the dressing method 
of Zakharov and Mikhailov \refs{\ZakharovPP,\ZakharovTY} for
constructing soliton solutions of classically integrable equations.
This is a very general technique, but we restrict our attention to
its application to the
principal chiral model,
since all of the solutions given in this paper may be embedded into it
as discussed below.

We consider a unitary $N \times N$ matrix field $g(z,\zbar)$ subject
to the equation of
motion
\eqn\pceom{
\pbar \left( \p g \, g^{-1} \right) + \p \left( \pbar g \, g^{-1} \right) = 0.
}
The dressing method allow us to
start with some
given solution $g$ of this equation and construct a new solution
$g'$ by
\eqn\gauge{
g \to g' = \chi g
}
for some appropriately chosen $\chi$.  If $\chi$ were just a constant
matrix, this would be an uninteresting unitary transformation, so
to generate physically distinct solutions we want $\chi$ to depend
on $z$ and $\zbar$.

\subsec{Construction}

The dressing method construction proceeds by introducing an auxiliary variable
$\lambda$ (called the spectral parameter, not to be confused
with the 't Hooft coupling in \charges) and considering the
system of equations
\eqn\aux{
i \pbar \Psi = {A \Psi \over 1 + \lambda}, \qquad
i \p \Psi = {B \Psi \over 1 - \lambda}
}
for three matrices $\Psi(\lambda)$, $A$, and $B$
(it is crucial that $A$ and $B$ are independent of $\lambda$).

The relation between \aux\ and \pceom\ is as follows.  If we have any
solution $g$ to \pceom, then we can take
\eqn\aandb{
A = i \pbar g\,g^{-1}, \qquad B = i \p g\,g^{-1}
}
and then solve \aux\ to find $\Psi(\lambda)$ such that
\eqn\initial{
\Psi(0) = g.
}
On the other hand, suppose we have any collection $(\Psi(\lambda), A, B)$
which satisfies \aux\ for all $\lambda$.  Then it is easy to check that
$\Psi(0)$ is guaranteed to satisfy \pceom.
We impose on $\Psi(\lambda)$ the unitarity condition
\eqn\conjugate{
\Psi^\dagger(\lambdabar) \Psi(\lambda) = 1.
}

Suppose we consider the analogue of the gauge transformation
\gauge\ for the auxiliary system \aux, now with a $\lambda$-dependent
gauge parameter
$\chi(\lambda)$,
\eqn\abprime{\eqalign{
\Psi  &\to \Psi' = \chi \Psi, \cr
A &\to A' = \chi A \chi^{-1} + i (1 + \lambda) \pbar \chi \chi^{-1}, \cr
B &\to B' = \chi B \chi^{-1} + i (1 - \lambda) \p \chi \chi^{-1}.
}}
If we can arrange for $\chi(\lambda)$ to be chosen in such a way
that the new $A'$ and $B'$ remain independent of $\lambda$, then
$(\Psi'(\lambda),A',B')$ is a legitimate new solution of
\aux, and hence provides a new solution $g' = \Psi'(0)$
of the principal chiral model.

The constraint that $A'$ and $B'$ should be
independent of $\lambda$ is easy to solve by imposing
constraints on the analytic properties of $\chi(\lambda)$
in the complex $\lambda$-plane.
Specifically, we require that $\chi(\lambda)$
should be meromorphic, and that
$\chi(\lambda) \to 1$ as $\lambda \to \infty$.
We say that $\chi(\lambda)$ has a pole at some $\lambda$ if
any entry of the matrix $\chi(\lambda)$ has a pole there.

Let us demonstrate by means of a simple example how these
analyticity constraints may be used to construct the desired $\chi(\lambda)$.
In the simplest nontrivial case, $\chi(\lambda)$ has a single pole at
some location $\lambda_1$.
In order to preserve the unitarity condition \conjugate,
$\chi(\lambda)$ should satisfy
\eqn\conjchi{
\chi^\dagger(\lambdabar)
\chi(\lambda) = 1.
}
Consequently $\chi^{-1}(\lambda)$ must have a
single pole at $\lambdabar_1$.
Already this information is enough to fix the dressing function $\chi$
to be  of the form
\eqn\blashke{
\chi(\lambda) = 1 + {\lambda_1 - \lambdabar_1 \over \lambda - \lambda_1} P
}
where $P$ is a hermitian projection operator (i.e., $P^2 = P = P^\dagger$).

It remains to choose $P$ so that $A'$ and $B'$
are independent of $\lambda$.
In fact, since they become independent of $\lambda$ as $\lambda \to \infty$,
it is sufficient to check that they have no poles.
Looking at \abprime\ we might worry
that they develop poles at $\lambda_1$ (from
$\chi(\lambda)$) or $\lambdabar_1$ (from $\chi^{-1}(\lambda)$).
It is however easy to check, using
the fact that $\Psi(\lambda)$ satisfies
the differential equations \aux, that the residues at these putative
poles actually vanish if one chooses the projection operator $P$ such
that its image is spanned by a collection of vectors of the form
$\{ \Psi(\lambdabar_1) e_1, \Psi(\lambdabar_1) e_2, \ldots \}$ where $e_i$ are
an arbitrary collection of constant vectors (independent of $z$ and
$\zbar$).  In general the projector $P$ can have any rank,
but in all of
our applications below $P$ will have rank one, so we write it
explicitly as
\eqn\projector{
P = {\Psi(\lambdabar_1) e e^\dagger \Psi^{-1}(\lambda_1) \over
e^\dagger \Psi^{-1}(\lambda_1) \Psi(\lambdabar_1) e}
}
for an arbitrary constant vector $e$.
It is clear that the overall scale of $e$ drops out of \projector,
so in fact $e$ lives in $\IP^{N-1}$, which parametrizes the
set of lines in $\IC^N$.
More generally, the data $\{e_i\}$ for a rank $k$ projection
operator would be specified by giving an element of the Grassmannian
Gr($k,N$) of $k$-planes in $\IC^N$.

\subsec{Summary for the U($N$) principal chiral model}

To summarize, the dressing method proceeds as follows.
Given any solution $g$ to the original equation \pceom, we first
solve the linear system \aux\ with $A$ and $B$ given by
\aandb\ to find $\Psi(\lambda)$.
The dressed solution $\Psi'(\lambda) = \chi(\lambda) \Psi(\lambda)$
may be constructed using
\blashke\ and \projector.
Finally, $g' = \Psi'(0)$ provides a new solution of \pceom.

It is clear that successive applications of this simple procedure,
i.e. $\Psi''(\lambda) = \chi'(\lambda) \Psi'(\lambda)$ etc.,
can be used to generate multi-soliton solutions.
We will illustrate this construction below via several examples.

\subsec{Reduction to the SO($N$) vector model}

Although the principal chiral model enjoys the most straightforward
application of the dressing method,
the equations \stringeom\ describing
conformal gauge
strings on $\IR \times S^{N-1}$
are those of the SO($N$) vector model.
Imposing the Virasoro constraints \virasoro\ gives the so-called
reduced \PohlmeyerNB\ vector model.
We can employ the dressing method for this model by
embedding it into the principal chiral model.

We choose the embedding following \refs{\SaintAubinAN,\HarnadWE,
\EichenherrHZ} (a different choice is shown in \OgielskiHV).
Define the diagonal $N \times N$ matrix
\eqn\thetadef{
\theta = {\rm diag}(+1,-1,\ldots,-1).
}
Then we choose the embedding of the vector $X_i$ into an SO($N$) principal
chiral field according
to the formula
\eqn\soembed{
\{ X_i : X_i X_i = 1 \} \qquad \leftrightarrow \qquad
g = \theta (2 X X^{\rm T} - 1) \in {\rm SO}(N).
}
Note that $g$ satisfies the identity
\eqn\coset{
g \theta g \theta = 1.
}
Geometrically, this identity specifies a particular coset
$S^{N-1} = {\rm SO}(N)/{\rm SO}(N-1)$ sitting inside SO($N$).
The dressing method proceeds as in the previous subsection, except
that we should add to \conjugate\ the additional conditions
\HarnadWE
\eqn\soconditions{
\overline{\Psi(\lambdabar)} = \Psi(\lambda), \qquad
\Psi(\lambda) = \Psi(0) \theta \Psi(1/\lambda) \theta.
}
In order to preserve \soconditions\ under dressing, the
dressing factor $\chi(\lambda)$ must satisfy
\eqn\conditions{
\overline{\chi(\lambdabar)} = \chi(\lambda), \qquad
\chi(\lambda) = \chi(0) \Psi(0) \theta \chi(1/\lambda) \Psi(0) \theta.
}
It is not possible for these constraints to be satisfied if
$\chi(\lambda)$ has a single pole.

Instead, there are two distinct classes of `minimal' solitons:
the simplest has two poles
\eqn\imagesone{
{\rm class~I}: \qquad \lambda_1,\quad \lambdabar_1 = 1/\lambda_1
}
located at conjugate points
on the unit circle,
while the second
has four poles located at
an arbitrary point $\lambda_1$ in the complex plane and its three images
under conjugation and inversion,
\eqn\images{
{\rm class~II}: \qquad \lambda_1, \quad
\lambdabar_1, \quad 1/\lambda_1, \quad 1/\lambdabar_1.
}
Below we will consider examples of both classes of solitons.
We will also consider the case of two class I solitons,
with two pairs of conjugate poles on the unit circle, being distinct
from a single class II soliton.

For class I the dressing factor is
\refs{\SaintAubinAN,\HarnadWE}
\eqn\classonechi{
\chi(\lambda) = 1 + {\lambda_1 - \lambdabar_1 \over \lambda - \lambda_1}
P + {\lambdabar_1 - \lambda_1 \over \lambda - \lambdabar_1} \overline{P}
}
with the projector $P$ given by the same formula \projector.
The constraints \conditions\ imply that the constant vector $e \in
\IC^N$
must satisfy
\eqn\classonee{
e^{\rm T} e = 0, \qquad \bar{e} = \theta e.
}

The construction of the dressing factor for class II solitons is
somewhat more complicated.
The reader can find all of the details in Theorem 4.2 and section 5 of
\HarnadWE.
In the example we look at below we will see that the class II soliton
with four poles \images\ in the complex plane
can be obtained from an analytic continuation of two pairs of
poles on the 
unit circle describing two class I solitons \imagesone.

\newsec{
Giant Magnons on $\IR \times S^3$ from the
U(2) Principal Chiral Model}

String theory on $\IR \times S^3$ admits a particularly simple
application of the dressing method since the string equations of
motion, in conformal gauge, are equivalent to those of the
SU(2) principal chiral model, via the embedding
\eqn\embed{
\{ (Z_1, Z_2) : |Z_1|^2 + |Z_2|^2 = 1 \}  \qquad \leftrightarrow \qquad
g = \pmatrix{Z_1 & -i Z_2 \cr - i \Zbar_2 & \Zbar_1} \in
{\rm SU(2)}.
}

One minor subtlety which arises for SU groups is that the dressing
factor \blashke\ does not have unit determinant.  Rather,
\eqn\detfactor{
\det \chi(\lambda) = {\lambda - \lambdabar_1 \over \lambda - \lambda_1}.
}
We can ensure that a dressed solution $\chi(0) \Psi(0)$
still sits in SU(2) (rather than U(2)) by throwing
in a compensating phase factor $(\lambdabar_1/\lambda_1)^{-1/2}$.

\subsec{The vacuum}

We begin with the vacuum solution
\eqn\vac{
\eqalign{
Z_1 &= e^{i t}, \cr
Z_2 &= 0
}
}
which describes a point-like string moving at the speed of light
around the equator of the $S^3$.  This state clearly has $\Delta - J = 0$.
Using the embedding \embed\ and \aandb\ we find
\eqn\aaa{
g_0 = \pmatrix{
e^{- i (z - \zbar)} & 0 \cr
0 & e^{+i (z - \zbar)}}, \qquad
A_0 = - B_0 = \pmatrix{-1 & 0 \cr
0 & 1}.
}
The corresponding vacuum solution $\Psi_0(\lambda)$
to the auxiliary problem \aux\ satisfying \initial\ is easily
found to be
\eqn\vacpsi{
\Psi_0(\lambda) = \pmatrix{
e^{+ i Z(\lambda)} & 0 \cr
0 & e^{-i Z(\lambda)}}, \qquad
Z(\lambda) = {z \over \lambda - 1} + {\zbar \over \lambda + 1}.
}

\subsec{A single two-charge soliton}

Let us now dress the vacuum \vacpsi\ to generate a one-soliton
solution.
We will show each step in great detail in order to demonstrate
the procedure clearly.
We  use the dressing factor \blashke\ with $P$ given
by \projector.  We can choose $e$ to be an arbitrary constant
element of $\IP^1$, which, without loss of generality, we can
parametrize as
\eqn\initiale{
e = (w, 1/w)
}
for $w \in \IC^*$.
Notice that $e$ only enters into \projector\ in the form
\eqn\shift{
\Psi_0(\lambdabar_1) e = \pmatrix{
w\,e^{+i Z(\lambdabar_1)} \cr
{1 \over w}\,e^{-i Z(\lambdabar_1)}}.
}
It is clear now that the complex parameter $w$ can be completely
absorbed by shifting $Z(\lambdabar_1) \to Z(\lambdabar_1) + i \log w$.
{}From \vacpsi\ it is evident that such a shift amounts to some particular
translation in the $x$ and $t$ coordinates.
Since this does not substantively affect the resulting solution,
we can without loss of generality go ahead and set $w=1$ for simplicity.

The projector $P$ can then be written as
\eqn\aaa{
P = {1 \over 1 + e^{2 i (Z(\lambda_1) - Z(\lambdabar_1))}}
\pmatrix{
1 & e^{+2 i Z(\lambda_1)} \cr
e^{-2 i Z(\lambdabar_1)} & e^{2 i (Z(\lambda_1) - Z(\lambdabar_1))}}.
}
The one-soliton solution is then
\eqn\psione{
\Psi_1(\lambda) = \left[ 1 + {\lambda_1 - \lambdabar_1 \over
\lambda - \lambda_1} P \right] \Psi_0(\lambda).
}
We can read off the corresponding solution in the $Z_i$ variables
from the embedding \embed, which leads to (keeping in mind the phase
discussed under \detfactor)
\eqn\solutionone{
\eqalign{
Z_1 &= {e^{+i t} \over |\lambda_1|}
{\lambda_1 e^{-2 i Z(\lambdabar_1)}
+ \lambdabar_1 e^{-2 i Z(\lambda_1)}
\over  e^{-2 i Z(\lambda_1)} + e^{-2 i Z(\lambdabar_1)}},\cr
Z_2 &=  {e^{-i t} \over |\lambda_1|}
{i(\lambdabar_1 - \lambda_1)
\over e^{-2 i Z(\lambda_1)} + e^{-2 i Z(\lambdabar_1)}}.
}}
One can check directly that this solves
the string equations of motion \stringeom, the Virasoro constraints
\virasoro, and satisfies the giant magnon boundary conditions
\magnondef.

It is instructive to express this solution in a more familiar form.
First we parametrize
\eqn\lambdadef{
\lambda_1 = r e^{i p/2}
}
and we introduce
\eqn\uvdef{\eqalign{
u &= i (Z(\lambda_1) - Z(\lambdabar_1)), \cr
v &= Z(\lambda_1) + Z(\lambdabar_1) - t,
}}
Plugging \lambdadef\ into \uvdef\ and using \vacpsi, we find that $u$
and $v$ may be expressed as
\eqn\uvexpres{
\eqalign{
u &=  \left[ x \cosh \theta  - t \sinh \theta \right] \cos \alpha,\cr
v &=  \left[ t \cosh \theta  - x \sinh \theta  \right] \sin \alpha,
}}
where $\alpha$ and $\theta$ are given by
\eqn\alphatheta{\eqalign{
\cot \alpha &= {2 r \over 1 - r^2} \sin {p \over 2}, \cr
\tanh \theta &= {2 r \over 1 + r^2} \cos {p \over 2}.
}}
Finally,
we find that the solution \solutionone\ may be written as
\eqn\aaa{\eqalign{
Z_1 &= e^{i t} \left[ \cos \hap + i \sin \hap \tanh u \right],\cr
Z_2 &= e^{i v} {\sin \hap \over \cosh u}.
}}
This form of the solution agrees precisely with the two-charge
giant magnon solution in \ChenGE, where it was shown to correspond
to the single-soliton solution of the complex sine-Gordon theory.
As a soliton of the U(2) principal chiral model, this solution
has been obtained in \ZakharovPP.
We also note that it reduces in the limit $r \to 1$ to the elementary
giant magnon of Hofman and Maldacena \HofmanXT.

If we force $p$ to lie within the range $-2 \pi < p < +2 \pi$,
then we see that
the total momentum \betterp\ is equal to $|p|$ for $-\pi < p < \pi$ and
$|p| - 2 \pi$ for $\pi < |p| < 2 \pi$.
In particular, $\lambda_1$ in the right half-plane gives
a soliton and $\lambda_1$ in the left half-plane
gives an anti-soliton.
The charges carried by this soliton may be obtained from \charges,
\eqn\aaa{\eqalign{
\Delta - J &= {\sqrt{\lambda} \over \pi}
{1 + r^2 \over 2 r} \left|\sin {p \over 2}\right|, \cr
J_2 &= {\sqrt{\lambda} \over \pi}
{1 - r^2 \over 2 r} \left|\sin {p \over 2}\right|.
}}
Eliminating $r$ between these two expressions gives the dispersion
relation \refs{\DoreyDQ,\ChenGE}
\eqn\aaa{
\Delta - J = \sqrt{J_2^2 + {\lambda \over \pi^2} \sin^2 {p \over 2}}.
}

\subsec{A scattering state of two two-charge solitons}

Now that we have all of the machinery set up, it is straightforward to
obtain the solution corresponding to two two-charge solitons.
We simply start with $\Psi_1(\lambda)$ given by \psione\ and
apply the dressing method a second time, now with
a pole at $\lambda = \lambda_2$.
In this manner
we obtain
\eqn\solutiontwo{\eqalign{
Z_1 &= {e^{i t} \over 2 |\lambda_1 \lambda_2|} {
R +
|\lambda_1|^2 \lambda_{1 \bar{1}} \lambda_{2 \bar{2}}
e^{+i (v_1 - v_2)}
+
|\lambda_2|^2 \lambda_{1 \bar{1}} \lambda_{2 \bar{2}}
e^{-i (v_1 - v_2)}
\over
\lambda_{12} \lambda_{\bar{1} \bar{2}} \cosh(u_1 + u_2)
+ \lambda_{1 \bar{2}} \lambda_{\bar{1} 2} \cosh(u_1 - u_2)
+ \lambda_{1 \bar{1}} \lambda_{2 \bar{2}} \cos(v_1 - v_2)},\cr
Z_2 &=  {-i \over 2 |\lambda_1 \lambda_2|}
{ \lambda_{1 \bar{1}} e^{i v_1} \left[
\lambda_{12} \lambda_{\bar{1} 2}  \lambdabar_2 e^{+u_2}
+ \lambda_{\bar{1} \bar{2}} \lambda_{1 \bar{2}} \lambda_2 e^{-u_2}
\right] + (1 \leftrightarrow 2) \over
\lambda_{12} \lambda_{\bar{1} \bar{2}} \cosh(u_1 + u_2)
+ \lambda_{1 \bar{2}} \lambda_{\bar{1} 2} \cosh(u_1 - u_2)
+ \lambda_{1 \bar{1}} \lambda_{2 \bar{2}} \cos(v_1 - v_2)},
}}
where
\eqn\aaa{
R = \lambda_{12} \lambda_{\bar{1} \bar{2}} \left[
\lambda_1 \lambda_2 e^{+u_1 + u_2} +
\lambdabar_1 \lambdabar_2 e^{-u_1 - u_2}
\right]
+ \lambda_{\bar{1} 2} \lambda_{1 \bar{2}} \left[
\lambda_1 \lambdabar_2 e^{+u_1 - u_2}
+ \lambdabar_1 \lambda_2 e^{-u_1 + u_2}
\right],
}
$u_i$ and $v_i$
are given by \uvdef\ with $\lambda_1 \to \lambda_i$, and we use the
shorthand notation
\eqn\aaa{
\lambda_{12} = \lambda_1 - \lambda_2, \qquad
\lambda_{1 \bar{2}} = \lambda_1 - \lambdabar_2, \qquad {\rm etc}.
}
Parametrizing $\lambda_i = r_i e^{i p_i/2}$, the conserved
charges of \solutiontwo\ are given by
\eqn\conservedtwo{\eqalign{
\Delta - J &= {\sqrt{\lambda} \over \pi}
{1 + r_1^2 \over 2 r_1} \left|\sin {p_1 \over 2}\right|
+ {\sqrt{\lambda} \over \pi}
{1 + r_2^2 \over 2 r_2} \left|\sin {p_2 \over 2}\right|, \cr
J_2 &= {\sqrt{\lambda} \over \pi}
{1 - r_1^2 \over 2 r_1} \left|\sin {p_1 \over 2}\right|
+ {\sqrt{\lambda} \over \pi}
{1 - r_2^2 \over 2 r_2} \left|\sin {p_2 \over 2}\right|.
}}
It is evident that \solutiontwo\ represents a scattering state
composed of two solitons of the type given in
\solutionone\ and discussed in \refs{\DoreyDQ,\ChenGE}.

It is interesting to note that \solutiontwo\ admits a
simple two (real) parameter generalization.
Recall that the construction of the projector $P$ in the dressing method
requires the choice \initiale\ of a vector $e$ which we parametrized
as $e = (w,1/w)$ for some non-zero complex number $w \equiv w_1$.
Previously, when
we had only a single soliton, we argued that $w_1$ could be set to 1
without loss of generality by a suitable translation of $x$ and $t$.
When applying the dressing method a second time to obtain the two-soliton
solution, we again have the freedom to choose a different arbitrary vector
$e_2 = (w_2,1/w_2)$, and there is no need for $w_1$ and $w_2$ to be
related.

It is still true that we can absorb $w_i$ into  $u_i$ and $v_i$
through \uvdef\ for $i=1,2$ separately.
The effect of this freedom is that \solutiontwo\ can
be generalized by taking
\eqn\freedom{
u_i \to u_i + a_i, \qquad v_i \to v_i + b_i, \qquad i=1,2
}
for four arbitrary real numbers $a_i$, $b_i$.
Two of these parameters can be absorbed by a suitable translation of
$x$ and $t$, but the remaining two parameters modify the shape of
the classical solution \solutiontwo\ nontrivially and therefore
correspond to `moduli' of the scattering state.

The solution \solutiontwo\ can also be mapped to a two-soliton solution
of the complex sine-Gordon theory by taking
the angular field $\phi$ of CSG
to be \DoreyDQ
\eqn\aaa{
\cos \phi = \pbar X_i \p X_i.
}
It would be interesting to see whether \solutiontwo\ could also
be obtained by exploiting the permutativity of the B\"acklund transformation
along the lines of \MikhailovZD.
Finally, it would be interesting to calculate the scattering
phase for \solutiontwo\ (in the string theory picture), along the lines
of \HofmanXT.

We have demonstrated how to apply the dressing method to the
problem of constructing superpositions of two-charge
solitons.  It is clear that this method can be used to generate
$n$-soliton scattering
solutions for any $n$, although the expressions are probably
too cumbersome to be of great use.
The generalizations of \conservedtwo\ and \freedom\ to arbitrary $n$
are obvious.

\subsec{A bound state of two two-charge solitons?}

It is also interesting to contemplate a bound state of two of these
two-charge solitons, along the same lines as the bound state of one-charge
solitons considered in \HofmanXT.
We begin by noting that in the solution \solutiontwo, $\lambda$
and $\lambdabar$ are completely free parameters. The expressions
given there, together with the corresponding $\Zbar_j$, which are obtained
by taking $i \to -i$ and exchanging $\lambda_j \leftrightarrow \lambdabar_j$,
satisfy the equations of motion \stringeom\ for arbitrary complex
values of $\lambda_j$ and $\lambdabar_j$.
In order for \solutiontwo\ to be a legitimate solution of the
$S^3$ sigma-model, however, we need to impose that the
sigma-model fields $X_i$ \complexZ\ are real.
This can be achieved by imposing, as we usually do,
that $\lambdabar_j$
is the complex conjugate of $\lambda_j$.
However this reality condition is also satisfied by taking
$\lambda_1$ to be the complex conjugate of $\lambdabar_2$ (and
vice versa),
a possibility that we now put to use.

For the bound state corresponding to a breather we are interested
in analytically continuing $p_i$ to complex momenta
\eqn\analytic{
p_1 = p + i q, \qquad p_2 = p - i q.
}
Using $\lambda_i = r e^{+i p_i/2}$ and $\lambdabar_i = r e^{-i p_i/2}$
we find
\eqn\complexlambdas{
\eqalign{
&\lambda_1 = r e^{-q/2} e^{+i p/2}, \qquad
\lambdabar_1 = r e^{+q/2} e^{-i p/2}, \cr
&\lambda_2 = r e^{+ q/2} e^{+i p/2}, \qquad
\lambdabar_2 = r e^{-q/2} e^{-i p/2},
}}
where we have already set $r_1 = r_2 = r$ to preserve the reality condition.
Surprisingly, the classical solution for these values of $\lambda_i$
and $\lambdabar_i$ is {\it identical} to the solution \solutiontwo\ for
$r_1 = r e^{-q/2}$, $r_2 = r e^{+q/2}$ and $p_i = p$.
Therefore the analytic continuation \analytic\ gives back
a scattering state rather than a bound state\foot{We are grateful
to J.~Maldacena for pointing this out to us.}.

Actually this result follows from the more general fact that
the solution \solutiontwo\ is completely symmetric under the exchange
$\lambdabar_1 \leftrightarrow \lambdabar_2$ with $\lambda_i$ held fixed.
This
has been pointed out in \DoreyXN, where the implications of this
fact for the singularity
structure of the magnon S-matrix
have been clarified.

In subsection 5.4 below
we will be able to use 
the analytic continuation \analytic\ 
to construct a true bound state of elementary (singly-charged)
Hofman-Maldacena giant magnons.

\newsec{
Giant Magnons on $\IR \times S^{N-1}$ from the
SO($N$) Vector Model}

In this section we apply the dressing method for the SO($N$) vector
model to giant magnon  solutions on $\IR \times S^{N-1}$.
Some of the solutions in this section may be obtained as limiting
cases of the $\IR \times S^3$ solutions
obtained in the previous section, but the method described here is
clearly more general since it can be applied to $\IR \times S^{N-1}$
for $N > 4$.

\subsec{The vacuum}

As before \vac, we start with the solution describing a point-like
string moving at the speed of light along the equator of the sphere,
\eqn\sovac{
X_i = (\cos t, \sin t, 0).
}
For the moment we work with SO($3$), describing strings on
$\IR \times S^2$.  The extension to SO($N$) is of course straightforward
and will be employed below.

We embed \sovac\ into the SO($3$) principal chiral model using
\soembed, and find the corresponding vacuum solution $\Psi_0(\lambda)$
to the linear system \aux\ is
\eqn\sovacpsi{
\Psi_0(\lambda) = \pmatrix{
\cos 2 Z(\lambda) & \sin 2 Z(\lambda) & 0 \cr
- \sin 2 Z(\lambda) & \cos 2 Z(\lambda) & 0 \cr
0 & 0 & 1},
}
where $Z(\lambda)$ is given as before by \vacpsi.
This form of
$\Psi_0(\lambda)$
has been chosen to satisfy the conditions \conjugate, \conditions.

\subsec{The HM giant magnon from a pair of poles on the unit circle}

The simplest soliton is obtained from the dressing factor
\classonechi\ and has two poles at conjugate points on the unit circle,
\eqn\aaa{
\lambda_1 = e^{+ i p/2}, \qquad \lambdabar_1 = e^{-i p/2}.
}
The projector $P$ is given by \projector, where we choose to parametrize
the initial vector $e$ as
\eqn\initialhme{
e = (1, i \sin w, i \cos w),
}
where $w$ is a real parameter.
Up to an overall (real) scale factor, which drops out of \projector\ anyway,
this is the most general choice satisfying the constraints \classonee.
It turns out that $w$ is an essentially irrelevant parameter and
may be absorbed into a translation of $x$ or $t$ (although the analysis which
leads to this conclusion is not quite as simple here as it was
in the case considered under \shift).
We can therefore set $w = 0$.

We use \initialhme\ and \sovacpsi\ to construct the projector $P$ shown
in \projector\ and the dressing factor $\chi(\lambda)$ given in
\classonechi.  Then $g = \chi(0) \Psi_0(0)$
is a solution of the SO(3) principal chiral model which lives
on the submanifold \coset, so that we can use \soembed\ to read off the new
solution $X$ in the $S^2$ sigma-model coordinates.  We find
\eqn\hmsol{
\eqalign{
X_1 + i X_2 &= e^{i t} \left[ \cos \hap + i \sin \hap \tanh u \right],\cr
X_3 &= \sin \hap \, {\rm sech}\, u
}
}
with $u$ given by
\eqn\udef{
u = \left[ x  - t \cos {p \over 2} \right] \csc {p \over 2}.
}
The dispersion relation is
\HofmanXT
\eqn\hmdisp{
\Delta - J = {\sqrt{\lambda} \over \pi} \left| \sin {p \over 2} \right|.
}
This is precisely the elementary giant magnon solution of
string theory on $\IR \times S^2$ found
by Hofman and Maldacena \HofmanXT, and the formula \hmdisp\ agrees
with the strong coupling limit of the exact magnon dispersion
relation \refs{\BeisertTM,\SantambrogioSB,\BerensteinJQtwo}
\eqn\aaa{
\Delta - J = \sqrt{1 + {\lambda \over \pi^2}
\sin^2{p \over 2}}.
}

The solution \hmsol\ also appears in
\SaintAubinAN\ as a solution of the
O(3) principal chiral model, and corresponds via the map between
strings on $\IR \times S^2$ and the sine-Gordon model to a single
sine-Gordon soliton.
Of course this solution may also be obtained by taking the $r \to 1$ limit
of \solutionone, in which the single pole moves onto the unit circle
and the charge $J_2$ goes to zero.

As mentioned above, one can check that the parameter $w$
which we set to zero
in \initialhme\ can be absorbed into a translation of $u$, which
in turn can be absorbed into a translation of $x$ or $t$.
The generalization of this elementary soliton from $\IR \times S^2$
to $\IR \times S^{N-1}$ is straightforward.  We parametrize
\initialhme\ by
\eqn\generalv{
e = (1, i \sin w, i \vec{v} \cos w)
}
where $\vec{v}$ is an arbitrary $N-2$ component unit vector.
As usual, the parameter $w$ can without loss of generality be set
to zero by an appropriate translation in $u$.
The unit vector $\vec{v}$ then specifies the orientation of the HM soliton
in the $N-2$ directions $(X_3,\ldots)$.

\subsec{A scattering state of two HM giant magnons from two pairs of
poles on the unit circle}

We can further dress the solution of the previous subsection by
adding a second pair of poles on the unit circle at $\lambda_2 =
e^{+ i p_2/2}$ and $\lambdabar_2 = e^{-i p_2/2}$.
This leads to the solution \juannote
\eqn\twohm{\eqalign{
X_1 + i X_2 &= e^{i t} + {e^{i t} (R + i I) \over
\sin\hapone \sin\haptwo (1 + \sinh u_1 \sinh u_2)
- (1 - \cos\hapone\cos\haptwo) \cosh u_1 \cosh u_2}
,\cr
X_3 &= {(\cos \hapone - \cos \haptwo) (\sin \hapone \cosh u_2
- \sin \haptwo \cosh u_1) \over
\sin\hapone \sin\haptwo (1 + \sinh u_1 \sinh u_2)
- (1 - \cos\hapone\cos\haptwo) \cosh u_1 \cosh u_2},
}}
with $u_i$ as in \udef, and
\eqn\aaa{\eqalign{
R &=
(\cos \hapone - \cos \haptwo)^2 \cosh u_1 \cosh u_2,\cr
I &= (\cos \hapone - \cos \haptwo) (\sin\hapone
\sinh u_1 \cosh u_2 - \sin\haptwo \cosh u_1 \sinh u_2)
}}
This is the explicit formula for the two-soliton scattering state
whose scattering phase was calculated in \HofmanXT\ (although
the precise form was not needed there because the phase shift
can easily be related to that of two solitons in the sine-Gordon model).
Again \twohm\ may be obtained by taking $r_1, r_2 \to 1$ in \solutiontwo.
As expected, the energy of \twohm\ is
\eqn\aaa{
\Delta - J = {\sqrt{\lambda} \over \pi}\left|\sin {p_1 \over 2}\right|
+ {\sqrt{\lambda} \over \pi}\left|\sin {p_2 \over 2}\right|.
}

Continued application of the dressing method may be used to construct
a scattering state with arbitrarily many solitons.
Each soliton can carry a different orientation in the
$N-2$ transverse directions by an appropriate choice of the initial vector
\generalv, and there is always the freedom to take $u_i \to u_i + a_i$ for
arbitrary real constants $a_i$.

\subsec{A bound state of two HM solitons from four poles in the complex plane}

We can also take four poles, at 
an arbitrary point $\lambda_1$ in the complex plane and its three images
\images, which we parametrize as
\eqn\aaa{
\lambda_1 = e^{+i (p + i q)/2}, \qquad
\lambdabar_1 = e^{-i (p - i q)/2}, \qquad
1/\lambda_1 = e^{-i (p + i q)/2}, \qquad
1/\lambdabar_1 = e^{+i (p - i q)/2}.
}
Following Theorem 4.2 and section 5 of \HarnadWE\ gives
the solution \juannote
\eqn\hmbreather{
\eqalign{
X_1 + i X_2 &= {e^{i t} }  {
\sinh^2 \haq \cosh^2(u + i \hap)
+ \sin^2 \hap \sin^2(v + i \haq)
\over  \sinh^2 \haq \cosh^2u  +  \sin^2 \hap \sin^2v}
\cr
X_3 &=
{\sin p \sinh^2 \haq \cosh u \cos v  -
\sin^2 \hap \sinh q \sinh u \sin v
\over \sinh^2 \haq \cosh^2u  + \sin^2 \hap \sin^2v}
}}
where
\eqn\hmbreatheruv{\eqalign{
u &= {2 \sin \hap \over \cosh q - \cos p}\left[
x \cosh{q \over 2} - t \cos{p \over 2}
\right],\cr
v &= {2 \sinh \haq \over \cosh q - \cos p} \left[
t \cosh{q \over 2} - x \cos{p \over 2}
\right].
}}
with dispersion relation
\eqn\aaa{
\Delta - J = {\sqrt{\lambda} \over \pi} \left|\sin{p \over 2}\right|
2 \cosh{q \over 2}.
}
As evident from this dispersion relation and the fact that
\hmbreather\ is periodic in $v$ (for fixed $u$), this solution
represents a bound state of two HM solitons.
This state was also discussed in \HofmanXT\ (though the full
solution \hmbreather\ was not presented).

We can also obtain the solution \hmbreather\ by analytically continuing
\twohm\ as follows.  We take
\eqn\aaa{
p_1 = p + i q, \qquad p_2 = p - i q
}
from which it follows that
\hmbreatheruv\ and \udef\ are related by
\eqn\aaa{
u = \ha (u_1 + u_2), \qquad v = {1 \over 2 i}(u_1 - u_2).
}
Making these substitutions in \twohm\ gives \hmbreather.

It is clear that successive applications of the dressing method,
together with appropriate analytic continuations as applied here,
can be used to construct the classical solution corresponding to
a bound state of any number of elementary giant magnons.
Moreover one can also construct scattering states in which various
combinations of elementary magnons and bound states participate.

\subsec{A three-charge giant magnon}

The previous few subsections have demonstrated the utility of the
dressing method, applied to the SO($N$) vector model, for constructing giant
magnons on $\IR \times S^{N - 1}$.   Many of the simplest examples
can be embedded inside $\IR \times S^3$
and may therefore be obtained as limits of the U(2) principal
chiral solutions we considered in the previous section.

Although the dressing method for the U(2) principal chiral model is
simpler, the advantage
of the SO($N$) vector model is its wider applicability to $\IR \times S^{N-1}$
for $N > 4$.  We leave a thorough analysis of the general case to
future work, and end here with a particularly simple example
of a three-spin giant magnon on $\IR \times S^5$.
The solution is given in complex coordinates by
\eqn\threecharge{\eqalign{
Z_1 &= e^{i t} {\cos\alpha_1 \tanh u_1
\tanh u_2 - \cos\alpha_2 \over
\cos\alpha_1 - \cos\alpha_2 \tanh u_1 \tanh u_2},
\cr
Z_2 &= e^{i v_1} { \sqrt{\cos^2\alpha_1 - \cos^2\alpha_2} \over
\cos\alpha_1 \cosh u_1
- \cos\alpha_2 \sinh u_1
\tanh u_2}, \cr
Z_3 &= e^{i v_2} { \sqrt{\cos^2\alpha_1 - \cos^2\alpha_2} \over
\cos\alpha_1 \cosh u_2
\coth u_1
- \cos\alpha_2 \sinh u_2},\cr
}}
where
\eqn\aaa{\eqalign{
u_1 &= x \cos\alpha_1, \qquad v_1 = t \sin\alpha_1,\cr
u_2 &= x \cos\alpha_2, \qquad v_2 = t \sin\alpha_2.
}}
This solution is valid for $\sin^2\alpha_1 < \sin^2\alpha_2$, which we can
assume without loss of generality.
As we have encountered before, the solution
\threecharge\ has a four real parameter generalization given by
\freedom.  As usual, two of those parameters can be absorbed into
shifts of $x$ and $t$.  In the particular case of \threecharge, a
third parameter can be absorbed into a rotation of $Z_2$ or $Z_3$
by a constant phase factor.
The net result of this analysis is that
\threecharge\ has a single physical modulus which adjusts the shape of
the solution.
This modulus may be taken to be $u_2 \to u_2 + {\rm constant}$.

The solution \threecharge\ carries charges
\eqn\aaa{
J_2 = {\sqrt{\lambda} \over \pi} {\sin\alpha_1 \over |\cos\alpha_1|}, \qquad
J_3 = {\sqrt{\lambda} \over \pi} {\sin\alpha_2 \over |\cos\alpha_2|}
}
and has energy
\eqn\aaa{
\Delta - J = {\sqrt{\lambda} \over \pi} \left(
{1 \over |\cos\alpha_1|} + {1 \over |\cos\alpha_2|}
\right).
}
Eliminating $\alpha_1$
and $\alpha_2$ gives the dispersion relation
\eqn\aaa{
\Delta - J = \sqrt{J_2^2 + {\lambda \over \pi^2}} +
\sqrt{J_3^2 + {\lambda \over \pi^2}}.
}
It is evident that this solution represents a scattering state
consisting of two superimposed two-charge
solitons \solutionone, one with momentum $p = \pi$ and the other
with momentum $p = - \pi$.
Since the total momentum is zero, this solution is compatible
with the form of the spinning string ansatz made in \ArutyunovUJ\ and
can be obtained directly by solving the equations of motion of
the Neumann integrable system.

\bigskip
\noindent
{\bf Acknowledgements}

We are grateful to A.~Neitzke, I.~Swanson and especially J.~Maldacena
for useful discussions and suggestions.
MS and AV acknowledge support from the U.S.~Department of Energy under
grant number DE-FG02-90ER40542,
and AV acknowledges support as the
William D.~Loughlin member of the IAS.
The research of MS is also supported by NSF grant PHY-0638520.
Any opinions, findings, and conclusions or recommendations
expressed in this material are those of the authors and do not
necessarily reflect the views of the National Science Foundation.

\listrefs
\end